\documentclass[aps,preprint,cite,groupedaddress,showpacs]{revtex4}
\usepackage{amsmath,amscd,amssymb,epsfig}
\begin{document}
\bibliographystyle{apsrev}

\title[Non-Markovian SR]{Non-Markovian Stochastic Resonance:\\
three state model of ion channel gating}
\author{Igor Goychuk and Peter H\"anggi}
\affiliation{Institute of Physics, University of Augsburg,
Universit\"atsstr. 1, D-86135 Augsburg, Germany}
\author{Jose L. Vega and Salvador Miret-Art\'es}
\affiliation{Instituto de Matem\'aticas
y F\'{\i}sica Fundamental (IMAFF), Consejo Superior
de Investigaciones Cient\'{\i}ficas,
Serrano 123, 28006 Madrid, Spain}

\date{\today}

\begin{abstract}
Stochastic Resonance in single voltage-dependent
ion channels is investigated within  a three state
non-Markovian modeling of the ion channel conformational dynamics.
In contrast to a two-state description
one assumes the presence of an additional closed state for the ion channel
which mimics  the manifold of voltage-independent closed
subconformations (inactivated ``state''). The
conformational transition into
the open state occurs through a domain of voltage-dependent closed
subconformations (closed ``state'').
At distinct variance with a standard two-state  or also three-state
Markovian approach, the inactivated state
is  characterized by a
broad, non-exponential probability distribution of corresponding
residence times. The linear response to a  periodic voltage signal
is determined for arbitrary distributions of the channel's recovery times.
Analytical results are obtained for the  spectral
amplification of the applied signal and the corresponding signal-to-noise ratio.
Alternatively, these results are also derived by use of a corresponding
two-state non-Markovian theory which is based on driven integral
renewal equations [I. Goychuk and P. H\"anggi,
Phys. Rev. E {\bf 69}, 021104 (2004)].
The non-Markovian features of stochastic resonance are
studied for a power law distribution of the residence time-intervals
in the inactivated state which  exhibits a large variance.
A comparison with the case of bi-exponentially distributed residence
times possessing the same mean value, i.e.  a
simplest non-Markovian  two-state description,  is also presented.

\end{abstract}

\maketitle

\section{Introduction}

Stochastic resonance (SR) \cite{review1,review2} is by now a well-established
phenomenon  with wide spread applications  in physics,
chemistry, engineering sciences and the life sciences. It refers to the fact that
in nonlinear stochastic systems an optimal level of applied or intrinsic noise
can dramatically boost the response (or, more generally,
the transport) to typically weak, time-dependent input signals.
This fact  plays a prominent role in biology with its abundance of a variety of
threshold-like systems that are subjected to
noise influences \cite{review1,review2,review3,review4}.
SR is often also characterized in terms of
an underlying stochastic synchronization between an applied
stimulus and an intrinsically
stochastic dynamics \cite{review1,review2,SRsynchro}.
From the viewpoint of physical biology, the phenomenon of SR
in biological sensory systems is commonly assumed  to be rooted
in the properties of excitable membranes. This being so,  it  ultimately can be
explained in terms of a driven stochastic dynamics of assemblies of ion channels
\cite{Bezrukov,Petracchi,Galvan,PRE00,PRE01,EPL01,EPLJung,FNL04}.
Although biologically relevant SR is generally  a property of a cooperative
coupling among ion channels \cite{EPL01,EPLJung,FNL04},
the study of stochastic resonance in single ion channels
carries merit  on its own: (1) SR has not yet been  convincingly
demonstrated on the level of {\it single} molecules, with biological
ion channels being such proper candidates \cite{Petracchi};
(2) ion channels can serve as suitable
single-molecular sensors to be utilized in  nanodevices.

All these facts in turn have stimulated a vivid interest on this subject matter.
Inspecting the recordings of ion current flowing through a single
ion channel, as obtained from  a typical patch
clamp experiment \cite{Hille,Neher}, bears close similarity with a
simple, driven bistable dynamics \cite{review1,review2,MNWJH}.
The occurrence of SR in single ion channels  thus seems very likely.
In reality, the situation is, however, more complex, because
(i) the dependence of the opening and closing rates
on the voltage and temperature are generally not Arrhenius-like
\cite{HodgkinHuxley,Marom1,PNAS02,PhysicaA2003}; a characteristic
feature which plays a central role for the
occurrence, or non-occurrence of the SR \cite{PRE00}, and (ii)
the detailed studies of the statistics of the
ion current switching events reveals that the probability distributions
of the residence time intervals in different conductance states are
normally not single exponential. This implies that the {\it observed} dynamics
of ``on-off'' conductance fluctuations is generically non-Markovian
within a two-state stochastic description
\cite{Liebovitch,West}. In the simplest nontrivial case of a bi-exponential
distribution of the residence time intervals dwelled in the
nonconducting conformation,
the emerging non-Markovian dynamics can be embedded into a  three state
Markovian dynamics \cite{Neher} with an extra
(visually indistinguishable) closed state.
The presence of such a third state (or, more generally, a number of
additional substates)
can, however,  be inferred
from the bursting character of the observed dynamics when
the ion channel dynamics  after switching between its open and closed states for a number of
times suddenly stops, and then  revives again later after a notably longer
time span has elapsed (as compared with the typical sojourn length in the open,
or closed state). The presence of such  a third state can be considered as a closed
inactivated state. This ``third'' state in fact does not constitute a single state, but
rather a manifold of many substates. As a consequence, the recovery of ion channel
from its inactive state does not present an exponential rate process, but will be governed
by a nonexponential distribution of corresponding residence times.
This feature, i.e. the absence of a well defined time scale for recovery from
inactivation can be observed in various ion channels \cite{Marom2}.

Our main objective here is  to
investigate the basic features of SR occurring in single ion channels
within the framework of  a three state non-Markovian dynamics.
In addition, we contrast this approach with the two-state
non-Markovian theory of stochastic resonance
developed in prior works \cite{PRL03,PRE04}.

\section{Model setup}

\begin{figure}
\begin{center}
\epsfig{file=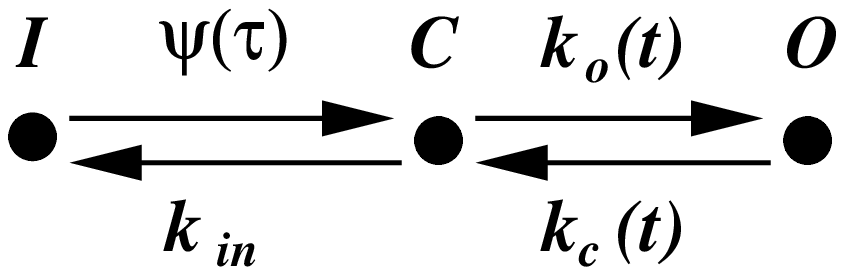,width=0.4\textwidth}
\end{center}
\caption{Sketch of the three state model setup with
generally time-dependent opening and closing rates and with a nonexponential
residence time statistics in the inactivated state.}
\label{Fig1}
\end{figure}

Following the reasoning put forward in Refs. \cite{PNAS02,PhysicaA2003}
we  consider a discrete state model of ion channel gating, whose essentials are depicted with
Fig. \ref{Fig1}. It consists of three ``states'': an open one ($O$), a closed one ($C$)
and an inactivated state ($I$). The inactivated state $I$ is assumed to correspond
to the manifold of voltage-insensitive conformations of the ion channel
protein and the opening
$C\to O$ and closing $O\to C$ transitions are associated with the motion of
a voltage sensor and the opening/closing of the channel's gate.
The stochastic transitions between the open and closed states are
characterized by generally time-dependent
opening  and closing  rates, $k_o(t)$ and $k_c(t)$, respectively.
These rates are assumed to depend on an externally
applied, time-dependent voltage signal.

In the spirit of the modeling put forward in
\cite{Millhauser,Condat,Bezanilla94,PNAS02,PhysicaA2003,GH04}
we assume that inactivation occurs from the closed state of the ion channel.
In doing so,  we are dealing with  an  archetype
model of gating with
opening, closing and inactivation dynamics \cite{Hille}. Furthermore,
unlike in the standard Markovian modeling \cite{Hille,Neher}
we assume  that the transition from the inactivated, voltage-independent
state to the closed state is not  rate-limited, but  rather is
characterized by a broad distribution of rates. Put differently, we model the
step from inactivation towards the closing state, cf. Fig. \ref{Fig1}
by a non-exponential distribution $\psi(\tau)$
of the residence time-intervals dwelled in the inactivated state.
This distribution
will be assumed to possess a finite
average $\langle \tau_r\rangle:=\int_0^{\infty}\tau\psi(\tau)d\tau$.
%%% add1
It is   voltage-independent but arbitrary otherwise. Furthermore,
it is assumed that the channel's inactivation occurs from
the closed state with a {\em voltage-independent} rate $k_{in}$.
%%% add2
The voltage-independence of $I\leftrightarrow C$ transitions
follows from the underlying character of conformational dynamics:
Namely, one assumes that the $I\leftrightarrow C$ transitions
occur in a direction being transverse to the direction of
$C\leftrightarrow O$ transitions \cite{PhysicaA2003}. These latter transitions
are tight to the
relocation of the gating charge across the membrane, while the former
transitions are not related to a charge redistribution; for further details we refer the
reader to the discussion in Ref. \cite{PhysicaA2003}.

This chosen distribution of rates accounts for the fact that
for several types of ion channels  the
recovery process does not have a well defined time
scale \cite{Marom2}. This circumstance in turn implies that the gating process is non-Markovian
within our three state description.  The standard Markovian three-state
description is recovered when $\psi(\tau)=k_r\exp(-k_r\tau)$, where
$k_r:=\langle \tau_r\rangle^{-1}$ is the rate
of the $I\to C$ transition.

\section{Non-Markov theory of three-state gating}

The time evolution for the probabilities to occupy the open,
closed or the inactivated
state, $p_o(t)$, $p_c(t)$ and $p_I(t)$, respectively, is governed
by the following generalized master equations:
\begin{eqnarray}\label{GME}
\dot p_o(t)& = & k_o(t)p_c(t)-k_c(t)p_o(t)\;, \nonumber \\
\dot p_c(t)& = &-(k_{in}+k_o(t))p_c(t)+k_c(t)p_o(t) +\int_{t_0}^{t}
\Gamma(t-t')p_I(t')dt'\;, \nonumber \\
\dot p_I(t) & = &k_{in} p_c(t) - \int_{t_0}^{t}
\Gamma(t-t')p_I(t')dt'\;.
\end{eqnarray}
The Laplace-transformed kernel $\Gamma(t)$ reads:
\begin{eqnarray}\label{kernel}
\tilde \Gamma(s)=\frac{s\tilde \psi(s)}{1-\tilde \psi(s)},
\end{eqnarray}
with $\tilde \psi(s)$ denoting the Laplace transform of $\psi(\tau)$
(``tilde'' denotes throughout the Laplace transform of a corresponding
function). The solution depends also on the initial condition,
chosen for example to read, $p_c(t_0)=1$, at initial time $t=t_0$.

This set of  equations can  formally  be derived following the approaches
in Refs. \cite{Kenkre73,BurshZharTem86,Hughes,
PRE04,Goychuk04}.
Alternatively, this very set of non-Markovian evolution eqs. can be obtained more directly as well:
The terms not expected from naive grounds in Eq.(\ref{GME})
are the ones that involve memory. Let us assume that the ion channel is prepared
in the state $I$, $p_I(0)=1$ at $t_0=0$ and impose $k_{in}\to 0$, i.e. no returns
are possible. Then, the leakage of probability $p_I(t)$ is due to
the transition into the state C, i.e.  $p_I(t)$ must equal (with such
an absorbing boundary condition)  the survival
probability $\Phi(t)=\int_t^{\infty}\psi(\tau)d\tau$. With this leakage given by
$\dot \Phi(t)=-\int_0^{t} \Gamma(t-t')\Phi(t')dt'$ we readily find  by use of
the Laplace transform method the
relation in Eq.(\ref{kernel}). The description in Eq.(\ref{GME}) portrays a
driven (i.e. inhomogeneous in time)
3-state , non-Markovian renewal process with the corresponding residence time
densities (RTDs) \cite{Cox,PRE04}, reading
\begin{eqnarray}
\psi_o(t+\tau,t)& = &k_c(t+\tau)\exp\left (-\int_{t}^{t+\tau}k_c(t')dt'
\right ),
\nonumber \\
\psi_c(t+\tau,t)& = &[k_{in}+k_o(t+\tau)]\exp\left(-\int_{t}^{t+\tau}[k_{in}+
k_o(t')]dt'\right), \\
\psi_I(\tau) & = &\psi(\tau). \nonumber
\end{eqnarray}
Note in particular that $\psi_o(t+\tau,t)$
and $\psi_c(t+\tau,t)$ depend not only on the length of residence time
intervals $\tau$, but also on the entrance time point $t$. We refer the reader to
Ref. \cite{PRE04}
for a detailed trajectory description of such driven renewal processes.
These conditioned, nonstationary two-time RTDs
$\psi_j(\tau|t):=\psi_j(t+\tau,t)$ ($j=o,c$) are given as the negative time
derivatives of the
corresponding survival probabilities $\Phi_{j}(\tau|t):=
\Phi_{j}(t+\tau,t)$; i.e.
$\psi_{j}(\tau|t):=-d \Phi_{j}(\tau|t)/d\tau$.

The averaged, time-dependent conductance $g(t)$ of the considered ion
channel reads $\langle g(t)\rangle=p_o(t)g_o+(p_c(t)+p_I(t))g_c$,
where $g_o$ and $g_c$ are the conductances of the open
and closed conformations, respectively. In the absence of a
time-dependent signal, $\langle g(t)\rangle=\langle g\rangle_{st}=
p_o^{st}g_o+(p_c^{st}+p_I^{st})g_c$, where $p_{\alpha}^{st}$
with $\alpha=o,c,I$ is the stationary solution of Eq. (\ref{GME})
in the absence of driving. For the following we assume a periodic
voltage signal given by
\begin{eqnarray}\label{signal}
V_s(t)=A\cos(\Omega t).
\end{eqnarray}
The mean deviation of the channel conductance $\langle \delta g(t)\rangle:=\langle g(t)\rangle-
\langle g\rangle_{st}$ from its stationary value
thus reads at asymptotic times within a linear response approximation \cite{review1},

\begin{eqnarray}\label{deviation}
\langle \delta g(t)\rangle=A|\tilde \chi(\Omega)|\cos(\Omega t-
\varphi(\Omega)) \;,
\end{eqnarray}
where $\varphi(\Omega)$ is the corresponding phase shift \cite{review1,GH04}.
In (\ref{deviation}),
$\tilde \chi(\Omega)$ denotes  the linear response function
in the frequency domain. The linear response result for the spectral amplification (SPA) of the signal
$\eta(\Omega)$  is given by
$\eta(\Omega)=|\tilde \chi(\Omega)|^2$ \cite{review1}.

\subsection{Linear response theory}
In order to evaluate the linear response function we
expand the solution of Eq.(\ref{GME}) as $p_c(t)=
\sum_{k=-\infty}^{\infty} r_k(t)\exp(-ik\Omega t)$ and $p_o(t)=
\sum_{k=-\infty}^{\infty} q_k(t)\exp(-ik\Omega t)$.
The solution for
$p_I(t)$ follows by virtue of  probability conservation, i.e.,
$p_I(t)=1-p_o(t)-p_c(t)$.
In the limit $t\to\infty$ the solutions become time-periodic \cite{review1, MNWJH}
with time-independent coefficients $r_k$ and
$q_k$ which depend on the amplitude strength $A$ and frequency $\Omega$.
The  asymptotic, nonlinear periodic
solution thus reads
\begin{eqnarray}\label{as}
p_c^{as}(t)& = &\sum_{k=-\infty}^{\infty} r_k\exp(-ik\Omega t);\;
r_{-k}=r_k^*, \nonumber \\
p_o^{as}(t)& = &\sum_{k=-\infty}^{\infty} q_k\exp(-ik\Omega t);\;
q_{-k}=q_k^*.
\end{eqnarray}
The linear response function
$\tilde \chi(\Omega)$ is encoded in the first Fourier term
\cite{GH04}. It reads,
\begin{eqnarray}\label{connect}
\tilde \chi(\Omega)=2\Delta g\lim_{A\to 0}\frac{q_1}{A} ,
\end{eqnarray}
 where $\Delta g:=g_o-g_c$ is the difference of ion
channel conductances in the
open and closed state, respectively.
%%%%%%%%%%%%%%%%%%%%%%%%%%%%%%%%%%%%%%%%%%%%%%%%

To start out, we assume the following small-signal expansion of
the time-dependent reaction rates
\begin{eqnarray}\label{small-rate}
k_{o,c}(t) & = &\nu_0\exp[-\Delta G_{o,c}(V(t))/k_BT]\nonumber \\
& \approx &
k_{o,c}^{(0)}(1-\beta_{o,c}V_s(t)),
\end{eqnarray}
where in absence of driving
\begin{eqnarray}\label{rate0}
k_{o,c}^{(0)}=\nu_0\exp[-\Delta G_{o,c}(V_0)/k_BT],
\end{eqnarray}
with $\Delta G_{o,c}(V_0)$ denoting the corresponding
static free energy barriers,
$V_0$ being a static voltage in the absence of signal,
i.e., $V(t) = V_0 + V_s (t)$, and
$\beta_{o,c}=-\frac{d\ln k^{(0)}_{o,c}}{dV}|_{V=V_0}$.

The substitution of  (\ref{as}) into (\ref{GME})
and taking (\ref{signal}) and (\ref{small-rate}) into account
yields for $t_0\to -\infty$ (this procedure
is equivalent to taking $t_0=0$ and $t\to\infty$)
the following recurrence relations:
\begin{eqnarray}\label{system}
[-ik\Omega+\tilde\Gamma(-ik\Omega)](r_k+q_k)=-k_{in}r_k+\frac{1}
{\langle \tau_r\rangle}\delta_{k,0}, \nonumber \\
(-ik\Omega+k_c^{(0)}) q_k+k_c^{(1)}(q_{k-1}+q_{k+1})=k_o^{(0)}r_k +
k_o^{(1)}(r_{k-1}+r_{k+1}),
\end{eqnarray}
where
\begin{equation}\label{k1}
k_{o,c}^{(1)}=-\frac{1}{2}A k_{o,c}^{(0)}\beta_{o,c}.
\end{equation}
In Eq. (\ref{system}), we used that $\lim_{s\to 0} \tilde
\Gamma(s)=1/\langle \tau_r\rangle$ which follows from
the expansion
$\tilde \psi(s)=1-s\langle \tau_r\rangle +o(s)$, where
$o(s)$ stands for the terms such that $\lim_{s\to 0}\frac{o(s)}{s}=0$.
From the first equation in (\ref{system})
we obtain
\begin{eqnarray}\label{gp0}
q_0=1-(1+k_{in}\langle \tau_r\rangle) r_0
\end{eqnarray}
for $k=0$ and
\begin{eqnarray}\label{gpk}
q_k=-\Big(1+\frac{k_{in}}{-ik\Omega+\tilde \Gamma(-ik\Omega)} \Big) r_k
\end{eqnarray}
otherwise.
Therefore, either $q_k$, or $r_k$ can be determined from the second equation
in Eq. (\ref{system}) which thus uncouples into a recurrence relation
for either $q_k$, or $r_k$. To the lowest order in $A$, $q_0=q_0^{(0)}+
O(A^2)$ and $r_0=r_0^{(0)}+ O(A^2)$ where $\{q_0^{(0)}, r_0^{(0)}\}$
are the values when the perturbation is absent.
A term linear in $A$ is absent
because a change from $A\to -A$ (which is equivalent
to a phase shift by $\pi$) cannot result into a shift of steady state
populations $q_0$ and $r_0$.  Moreover, $q_{1}\propto A$
and $r_{1}\propto A$ to leading order in $A$.
Likewise, the expansions of
 $q_{2}$ and $r_{2}$ in $A$ start out
 from $A^2$.

Then, by taking into account $r_0=k_c^{(0)}q_0/k_o^{(0)}+O(A^2)$ in Eqs.
(\ref{system}), (\ref{gp0}), (\ref{gpk}) we obtain after some algebraic manipulations
\begin{eqnarray}
r_0 & = & \frac{k_c^{(0)}}{k_o^{(0)}+
k_c^{(0)}(1+k_{in}\langle \tau_r\rangle)} + O(A^2)\;, \nonumber\\
q_0 & = &  \frac{k_o^{(0)}}{k_o^{(0)}+
k_c^{(0)}(1+k_{in}\langle \tau_r\rangle)}+ O(A^2)\;,
\end{eqnarray}
and
\begin{eqnarray}\label{main}
q_1=\frac{1}{2}A\frac{\beta_c-\beta_o}{\langle \tau_o\rangle+
\langle \tau_{c'}\rangle}\frac{1}{k_c^{(0)}-i\Omega+
\frac{i\Omega k_o^{(0)}}{i\Omega -k_{in}[1-\tilde \psi(-i\Omega)]}} + O(A^2),
\end{eqnarray}
where
\begin{eqnarray}
\langle \tau_{o}\rangle=1/k_c^{(0)}
\label{tauO}
\end{eqnarray}
is the mean residence time in the open state. Furthermore,
\begin{eqnarray}\label{tau}
\langle \tau_{c'}\rangle=(1+k_{in}\langle \tau_r\rangle)/k_o^{(0)}
\end{eqnarray}
is the mean residence time within the set of closed states $c'=(C,I)$
in the absence of time-dependent driving.

This latter quantity is defined
as $\langle \tau_{c'}\rangle:=\int_0^{\infty}\tau
\psi_{c'}(\tau)d\tau$, where $\psi_{c'}(\tau)$ is the stationary RTD in the set of
compound closed states, when no time-dependent signal acts, i.e. $V_s(t)=0$.
This auxiliary quantity is obtained as follows: The channel is prepared in the closed
state $C$ at $t_0=0$, i.e., $p_c(0)=1$ and the back transition $O\to C$
is set to zero by imposing $k_c\to 0$. Then, the solution of the
first two equations in (\ref{GME}) in the absence of driving
yields the stationary survival probability of the
compound closed states as $\Phi_{c'}(t)=p_c(t)+p_I(t)$ and $\psi_{c'}(t)$
follows as $\psi_{c'}(\tau)=-d\Phi_{c'}(\tau)/d\tau$. Using this scheme we find,
\begin{eqnarray}\label{compound0}
\tilde \Phi_{c'}(s)= \frac{1+\frac{k_{in}}{s+\tilde \Gamma(s)}}{s+k_o^{(0)}+
k_{in}\frac{s}{s+\tilde\Gamma(s)}}.
\end{eqnarray}
and in virtue of  $\tilde \psi_{c'}(s)=1-s\tilde \Phi_{c'}(s)$
\begin{eqnarray}\label{compound1}
\tilde \psi_{c'}(s)= \frac{k_o^{(0)}}{s+k_o^{(0)}+
k_{in}\frac{s}{s+\tilde\Gamma(s)}}\;.
\end{eqnarray}
By use of Eq. (\ref{kernel}) in Eq. (\ref{compound1})
this stationary distribution of the residence times can be recast as
\begin{eqnarray}\label{compound2}
\tilde \psi_{c'}(s)= \frac{k_o^{(0)}}{s+k_o^{(0)}+
k_{in}(1-\tilde \psi(s))}\;.
\end{eqnarray}
Furthermore,  $\langle \tau_{c'}\rangle$ in Eq. (\ref{tau})
follows from Eq. (\ref{compound0}) as
$\langle \tau_{c'}\rangle=\tilde \Phi_{c'}(0)$.

Let us introduce also the auxiliary function
\begin{eqnarray}\label{aux}
\tilde{ \mathcal{G}} (s) =\frac{(1-\tilde \psi_o(s))(1-\tilde \psi_{c'}(s))}
{1-\tilde \psi_o(s)\tilde \psi_{c'}(s)},
\end{eqnarray}
where $\tilde \psi_o(s)=k_c^{(0)}/(s+k_c^{(0)})$ is the Laplace transform of the stationary
RTD of the open time intervals. Then, Eq. (\ref{main}) can be
rewritten in a more compact form as

\begin{eqnarray}\label{main2}
q_1=\frac{1}{2}A\frac{\beta_c-\beta_o}{\langle \tau_o\rangle+
\langle \tau_{c'}\rangle}\frac{\tilde{\mathcal{ G}}(-i\Omega)}{-i\Omega} + O(A^2)\;.
\end{eqnarray}
Since $k_{in}\langle \tau_r\rangle$ does not depend
on voltage $V$,
$\beta_{o'}:=\frac{d\ln \langle \tau_{c'}\rangle}{dV}|_{V=V_0}=
-\frac{d\ln k^{(0)}_{o}}{dV}|_{V=V_0}=\beta_o$ and
the result in
Eq. (\ref{main2}) coincides with the result of the phenomenological
two-state theory
of non-Markovian SR put forward in \cite{PRL03,PRE04}.
Upon combining  with Eq. (\ref{connect}), Eq. (\ref{main2}) yields the linear response result
\begin{eqnarray}\label{PHLR}
\tilde \chi(\Omega)= \frac{(\beta_c-\beta_o)\Delta g}{\langle \tau_o\rangle+
\langle \tau_{c'}\rangle}\frac{\tilde{\mathcal{ G}}(-i\Omega)}{-i\Omega} \;,
\end{eqnarray}
which in fact coincides with Eq. (61) in Ref. \cite{PRE04}.

The just outlined procedure can as well be extended into the
nonlinear response
regime to obtain nonlinear response functions of  required
order in the signal amplitude.
This, however, is beyond the scope of this work.
This central result in Eq.(\ref{PHLR}) is alternatively derived
in Appendix A by making use of the two-state non-Markovian theory detailed in Ref.
\cite{GH04}.

\subsection{Explicit results in terms of thermodynamic  free energies}

We next introduce {\it formally} the effective free energy bias
$\epsilon(T)$, i.e.,
\begin{eqnarray}\label{def}
\frac{\langle \tau_{o}\rangle}{\langle \tau_{c'}\rangle}=
\exp\left(- \frac{\epsilon(T)}{k_BT} \right).
\end{eqnarray}
In accordance with the relations (\ref{tauO}, \ref{tau}) we obtain
\begin{eqnarray}\label{eps}
\epsilon(T)=\Delta G_o(V_0)-\Delta G_c(V_0)+k_B T\ln(1+k_{in}
\langle \tau_r\rangle)\;.
\end{eqnarray}
Using that $\Delta G_o(V_0)=G^{\#}-G_c$ and $\Delta G_c(V_0)=G^{\#}-G_o$,
where $G^{\#}$ is the free energy of the transition state, and
$G_{o,c}=H_{o,c}-TS_{o,c}$ is the free energy of the
open (closed) conformation, Eq.(\ref{eps}) can be recast as

\begin{eqnarray}\label{eps1}
\epsilon(T)=\Delta H- T\Delta S+k_B T\ln(1+k_{in}
\langle \tau_r\rangle)\;,
\end{eqnarray}
where $\Delta H:=H_o-H_c$ is the difference of thermodynamic enthalpies
of the open and closed
conformations
(or of the internal energies, if no volume change of the macromolecule
occurs at the conformational transition)
and $\Delta S:=S_o-S_c$ is the corresponding entropy
difference.

Assuming that the free energy barriers in Eq. (\ref{small-rate}) have linear
dependence of voltage, $\Delta G_o(V)=\Delta G_o-q(1-\delta)V$
and  $\Delta G_c(V)=\Delta G_c+q\delta\; V$,
where $q$ is the gating charge and $0<\delta<1$ is a constant
\cite{Hille,PhysicaA2003},
we find that $\beta_c-\beta_o=-q/(k_BT)$.
For the spectral amplification
of the conductance response $\eta=|\tilde \chi(\Omega)|^2$
we obtain then from Eq. (\ref{PHLR}) our first main result
\begin{equation}\label{eta}
\eta(\Omega,T)=
\frac{(\Delta g)^2 q^2}{16(k_BT)^2}
\frac{\nu^2(T)}{\cosh^4\left[\frac{\epsilon(T)}{2k_BT}\right]}
\frac{|\tilde{\mathcal{ G}}(-i\Omega)|^2}{\Omega^2}\;,
\end{equation}
with $\epsilon(T)$ given in Eq. (\ref{eps}), while
$\nu(T)=\langle \tau_o\rangle^{-1} + \langle
\tau_{c'}\rangle^{-1}$ denotes  the sum of the closing and opening
rates. $\tilde{\mathcal{ G}}(s)$ in
Eqs. (\ref{aux}),(\ref{eta}) reads explicitly
\begin{eqnarray}\label{Gfin}
\tilde{\mathcal{ G}}(s)=\frac{s}{s+k_c^{(0)}+\frac{s k_o^{(0)}}{s+k_{in}(1-
\tilde \psi(s))}} \;,
\end{eqnarray}
where $k_c^{(0)},k_o^{(0)},k_{in}$ are the unperturbed rates
and $\psi(\tau)$ is the residence time distribution.
The result in Eq.(\ref{eta})  coincides, apart from a model
specific constant,  with the expression given in Ref. \cite{PRL03},
cf. Eq.(25) therein. Within the
linear response approximation the corresponding
signal-to-noise ratio (SNR) at the angular driving frequency $\Omega$ is obtained from
$R_{SN}(\Omega,T):=\pi A^2\eta/S_N(\omega=\Omega)$, where
$S_N(\omega)$ denotes the spectral power density of conductance fluctuations
in the absence of signal $V_s(t)$. By use of the Stratonovich formula for
the autocorrelation function of the alternating renewal process
\cite{Strat} and the Wiener-Khinchin theorem  one finds for $S_N(\omega)$
\cite{Strat,PRE04}

\begin{equation}\label{spower}
S_N(\omega)=\frac{2(\Delta g)^2}{\langle \tau_o\rangle +\langle
\tau_{c'}\rangle}\frac{1}{\omega^2}
{\rm Re}\left [\tilde{\mathcal{ G}}(i\omega)\right]
\end{equation}
Therefore, the SNR equals the result derived in Refs.
\cite{PRL03,PRE04}, namely,
\begin{eqnarray}\label{res3}
{\rm R_{SN}}(\Omega,T)=\frac{\pi A^2q^2}{8(k_BT)^2}\frac{\nu(T)}
{\cosh^2\left[\frac{\epsilon(T)}{2k_BT}\right]}\; N(\Omega),
\end{eqnarray}
where
\begin{eqnarray}\label{N}
N(\Omega)=\frac{
|\tilde{\mathcal{ G}}(i\Omega)|^2}{{\rm Re}[\tilde{\mathcal{ G}}(i\Omega)]}
\end{eqnarray}
provides the specific function which accounts for manifest non-Markovian effects.
In the low frequency limit $\Omega\to 0$,
$N(\Omega)$ approaches the limit $N(0)={2}/({C_o^2+C_{c'}^2})$
\cite{PRL03,PRE04},
where $C_o$ and $C_{c'}$ are the coefficients of variation of the RTDs
of the open and compound closed states, respectively.

By use of the expansion $\tilde \psi(s)=1-s\langle
\tau_r\rangle+o(s)$ in Eq. (\ref{Gfin}), and Eq. (\ref{tau}) we find
in the adiabatic limit $\Omega\to 0$ that
the spectral power amplification  acquires the universal form, reading
\begin{eqnarray}\label{universal}
\eta(\Omega\to 0,T)=
\frac{(\Delta g)^2 q^2}{16(k_BT)^2}
\frac{1}{\cosh^4\left[\frac{\epsilon(T)}{2k_BT}\right]}\;.
\end{eqnarray}
This result holds in the presence of asymmetry with nonvanishing $\epsilon(T)$.
In such a case, the SPA (\ref{universal})
can exhibit a sharp stochastic resonance at the physiologically
relevant temperatures when $\epsilon(T)$ changes  sign, when the probabilities
of the channel to be open, or to stay
close become equal. This corresponds to the opening threshold for the
detailed below
case of an ion channel sensitive to cold (or, vice versa, sensitive to heat,
if the open state is preferred from an entropic viewpoint). The cold-sensitive,
or heat-sensitive ion channels \cite{Nature04} present appropriate candidates
to reveal this entropic-SR effect.
This feature  at very small driving frequencies, which mimics a Markovian behavior,
has thus nothing to do with non-Markovian properties; it is solely
due to an entropic asymmetry. The non-Markovian effects
emerge, however, at small, but finite frequencies in the corresponding SPA-curves,
see in Ref.\cite{PRL03}. Most prominent non-Markovian, long-time
memory
effects distinctly suppress, however, the SNR in the low-frequency domain \cite{PRL03,PRE04}.
For $\Omega\to 0$, we obtain (with $C_o=1$)
\begin{eqnarray}\label{SNR0}
{\rm R_{SN}}(\Omega\to 0,T)=\frac{\pi A^2q^2}{8(k_BT)^2}\frac{\nu(T)}
{\cosh^2\left[\frac{\epsilon(T)}{2k_BT}\right]}\frac{2}{1+C_{c'}^2},
\end{eqnarray}
with $C_{c'}>1$. Thus, the $R_{SN}(\Omega=0,T)$ is fully suppressed,
making the detection of low-frequency signals barely possible, e.g.,
for the
power law distribution $\psi(\tau)\propto 1/\tau^{2+\alpha} $ with
$0<\alpha<1$ considered below. This suppression occurs
due to  $1/f^{1-\alpha}$ noise feature in the spectral power
$S_N(\Omega)$ for small frequencies. Let us illustrate now these
general considerations with two particular models.

\section{Ion channel gating: Bi-exponential distribution versus a power law}

The current three state model provides a  suitable framework to clarify
the role of a power-law distributed residence times in the inactivated and closed states
as compared with the simplest two-state non-Markovian situation (with respect
to the {\it observable} dynamics) of a bi-exponential distribution,
that, likewise, can be embedded into a three-state Markovian description.

\subsection{The case of a bi-exponential distribution}

We start our driven channel gating investigation with the simplest case of an exponential
residence time distribution of the transition $I \to C$, i.e., $\psi(\tau)=k_r\exp(-k_r\tau)$.
Then, Eq. (\ref{compound2}) yields
\begin{eqnarray}\label{compound3}
\tilde \psi_{c'}(s)= \frac{k_o^{(0)}(s+k_r)}{(s+k_o^{(0)})(s+k_r)+
k_{in}s}\;.
\end{eqnarray}
The inversion of Eq. (\ref{compound3}) yields a bi-exponential
probability density, i.e.,
\begin{eqnarray}\label{biexp}
 \psi_{c'}(\tau)= c_1\lambda_1\exp(-\lambda_1\tau)+
 c_2\lambda_2\exp(-\lambda_2\tau)\;
\end{eqnarray}
with the rate coefficients $\lambda_{1,2}=\frac{1}{2}
(k_o^{(0)}+k_{in}+k_r\pm \sqrt{(k_o^{(0)}-k_{in}-k_r)^2
+4k_o^{(0)}k_{in}})$ and  corresponding weight factors $c_{1,2}=\frac{1}{2}
\left (1\pm (k_o^{(0)}-k_{in}-k_r)/\sqrt{(k_o^{(0)}-k_{in}-k_r)^2
+4k_o^{(0)}k_{in}}\right)$. The mean residence time corresponding to
this RTD (\ref{biexp}) is
\begin{eqnarray}\label{meanRTD1}
\langle \tau_{c'}\rangle =\frac{1}{k_o^{(0)}}\left (1+ \frac{k_{in}}{k_r}\right )
\end{eqnarray}
Such a  non-exponential, i.e. bi-exponential RTD can
actually be very broad as characterized
by the corresponding coefficient of variation $C:=\sqrt{
\langle \tau^2\rangle-\langle \tau\rangle^2}/\langle \tau\rangle$, yielding
from (\ref{biexp})
\begin{eqnarray}\label{CV1}
C_{c'}=\sqrt{1+\frac{2k_o^{(0)}k_{in}}{(k_{in}+k_r)^2}} \, .
\end{eqnarray}
Indeed, for $k_r \ll k_{in}\ll k_o^{(0)}$ we deduce from Eq.
(\ref{CV1}) that $C_{c'}\approx \sqrt{2k_o^{(0)}/k_{in}}\gg 1$.
This implies that the distribution (\ref{biexp}) has a  small, but
very broad long-time tail, which in turn results in a large variance of the residence
times. This finding carries important consequences: As shown in Refs.
\cite{PRL03,PRE04}
the signal-to-noise ratio $R_{SN}$
is then strongly suppressed in the low-frequency limit $\Omega\to 0$ by
the factor $1/N(0)=\frac{1}{2}(C_o^2+C_{c'}^2)$, where $C_o$ is the
coefficient of variation for $\psi_o(\tau)$, $C_o=1$ in the present case.
 This presents a first manifest non-Markovian effect which is present
already within this simplest non-Markovian setting. We also note that the
auxiliary function $\tilde{\mathcal{ G}}(s)$ can be recast as
\begin{eqnarray}\label{aux2}
\tilde{\mathcal{ G}}(s) =\frac{s(s+k_{in}+k_r)}{[s+\mu_1(T)][s+\mu_2(T)]},
\end{eqnarray}
where
\begin{eqnarray}
\mu_{1,2}(T) =\frac{1}{2}\left(k_o^{(0)} +k_c^{(0)}+k_{in}+k_r\pm
\sqrt{(k_o^{(0)} +k_c^{(0)}-k_{in}-k_r)^2+4k_o^{(0)}k_{in}} \right)
\end{eqnarray}
are the decay rates of the conductance time-correlations.
By use of Eqs. (\ref{aux2}), (\ref{N}) the results in Eqs. (\ref{eta}) for the SPA, and (\ref{res3})
for the SNR assume explicitly the form:
\begin{equation}\label{eta2}
\eta(\Omega,T)=
\frac{(\Delta g)^2 q^2}{16(k_BT)^2}
\frac{\nu^2(T)}{\cosh^4\left[\frac{\epsilon(T)}{2k_BT}\right]}
\frac{\Omega^2+(k_{in}+k_r)^2}
{[\Omega^2+\mu_1^2(T)][\Omega^2+\mu_2^2(T)]},
\end{equation}
and
\begin{eqnarray}\label{SNR2}
{\rm R_{SN}}(\Omega,T)=\frac{\pi A^2q^2}{8(k_BT)^2}\frac{\nu(T)}
{\cosh^2\left[\frac{\epsilon(T)}{2k_BT}\right]}
\frac{\Omega^2+(k_{in}+k_r)^2}{\Omega^2+(k_{in}+k_r)^2+k_o^{(0)}k_{in}}\; ,
\end{eqnarray}
respectively. The remaining parameters are: $\nu(T)=\langle \tau_o\rangle^{-1} + \langle
\tau_{c'}\rangle^{-1}$\/$=k_c^{(0)}+k_o^{(0)}k_r/(k_r+k_{in})$ with
$k_{o,c}^{(0)}$ in Eq. (\ref{rate0}) and $\epsilon(T)=
\Delta G_o(V_0)-\Delta G_c(V_0)+k_B T\ln(1+k_{in}/
k_r)$ with exponential Arrhenius rates $k_{in}=\nu_0\exp(-\Delta G_{in}/k_BT)$
and $k_r=\nu_0\exp[-\Delta G_r/k_BT]$ \cite{HTB90}.

The results in Eq.(\ref{eta2}) and (\ref{SNR2}) constitute  central results
for the SR occurring in a three state Markovian model
of gating in ion channels possessing an inactivation from the closed state.
At the same time, these results correspond to a simplest non-Markovian
two-state model of the observable dynamics of conductance fluctuations.

We performed  numerical calculations for the
set of test parameters given in the Table \ref{table} which is chosen
to mimic the experimental
temperature dependence of
the cold-sensitive ion channels, see in Ref.  \cite{Nature04}.
%%%%%%%%%%%%%%%% table %%%%%%%%%%%%%%
\begin{table}%[H] add [H] placement to break table across pages
\caption{Free energy barriers,
$\Delta G_\alpha =\Delta H_{\alpha}-T\Delta S_{\alpha}$\label{table}}
\begin{ruledtabular}
\begin{tabular}{lcr}
$\alpha$ & Enthalpy part $\Delta H_{\alpha}$, {\rm kJ/mol} \footnote{
Table depicts the corresponding parameters for the 
free energy barriers that enter the related rate
coefficients. Here, $\Delta H_{\alpha}$ must be
divided with the Avogadro number $N_A$ to obtain the corresponding value
per single molecule. For example, $\Delta H_c=175\; {\rm kJ/mol}$ 
thus corresponds  (we
use the Boltzmann constant $k_B$ and not the gas constant
$R=k_B N_A$ in the rate expressions, respectively) to a value,
$\Delta H_c\approx 2.91\cdot 10^{-19}\;J\approx 1.81\;{\rm eV}$, etc.}
& Entropy part $\Delta S_{\alpha}$, $k_B$ \\
\hline
$c$ & 175 & 52\\
$o$ & 15  & -10 \\
$in$ & 15 & -15 \\
$r$  & 15 & -20 \\
% Lines of table here ending with \\
\end{tabular}
\end{ruledtabular}
\end{table}
%%%%%%%%%%%%%%%%%%%%%%%%%%%%%%%%%

\begin{figure}
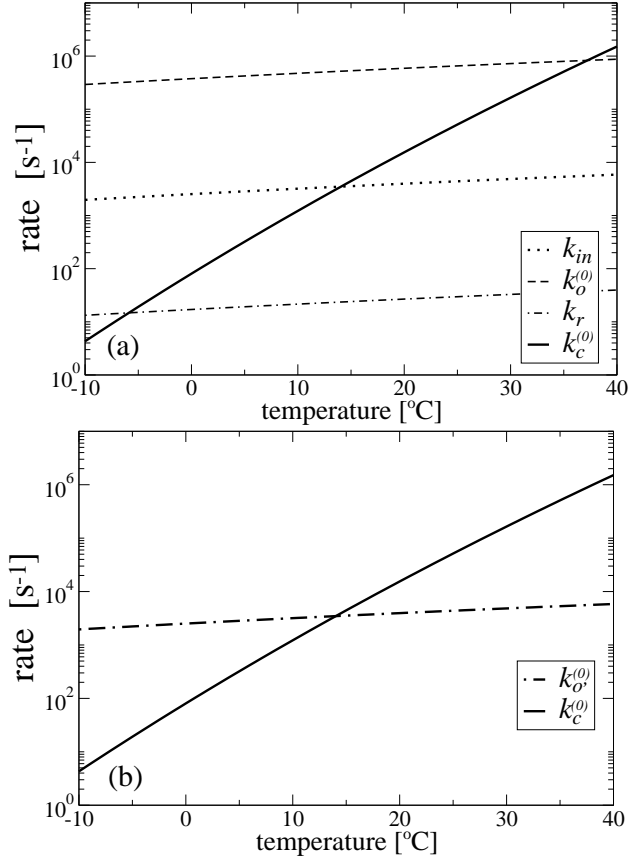

\epsfig{file=Fig2nm3a.eps,width=.5\textwidth}
\vspace{1cm}
\epsfig{file=Fig2nm3b.eps,width=.5\textwidth}
\vspace{1cm}
\caption{Temperature-dependence of the rates used for the ion
channel gating dynamics, absent the driving.
Part (a) for the considered three state
Markovian model; part (b) for the corresponding two-state non-Markov 
model with the effective opening
rate defined as $k_{o'}^{(0)}:=\langle \tau_{c'}\rangle^{-1}$,
where $\langle \tau_{c'}\rangle$ is the mean residence time.
The rate $k_{c}^{(0)}$ denotes again the transition from the open state $O$
towards the closed state $C$ in the absence of time-dependent driving.
}
\label{Fig2}
\end{figure}

\begin{figure}
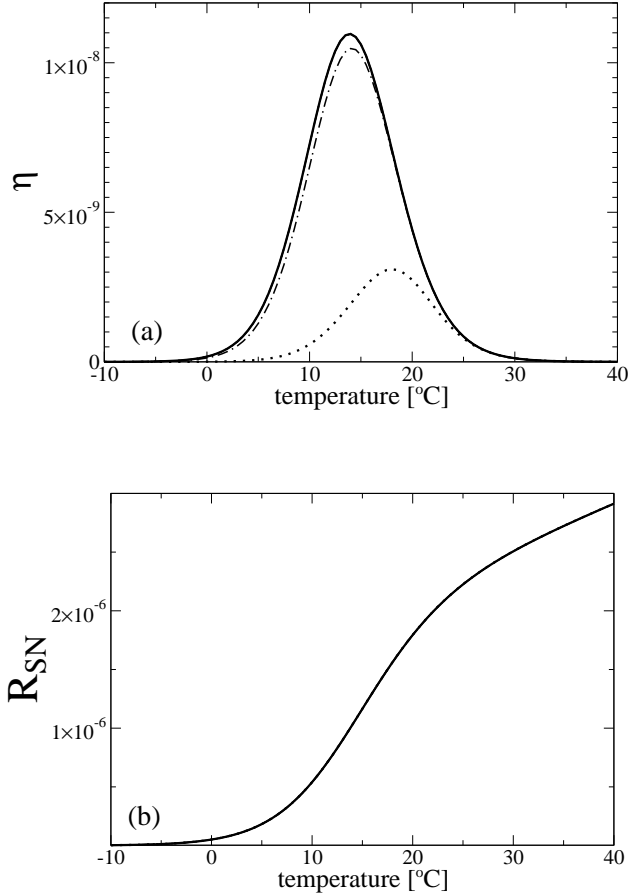

\epsfig{file=Fig3nm3a.eps,width=.5\textwidth}
%\hfill

\vspace{1cm}

\epsfig{file=Fig3nm3b.eps,width=.5\textwidth}
\vspace{1cm}
\caption{Markovian three-state model of ion channel gating: (a) spectral power amplification $\eta$
(in units of $(\Delta g q)^2/(J/mol)^2$) and (b) signal to  noise
ratio $R_{SN}$ (in units of $\pi(qA)^2/(J/mol)^2\cdot {\rm s}^{-1}$)  versus
temperature
(in $^o C$) at different angular frequencies $\Omega$ of the  harmonic signal.
In Fig.(a),  the full line corresponds to the adiabatic limit
$\Omega\to 0$; the dash-dotted line corresponds to
$\Omega=10\;{\rm s}^{-1}$ and the dotted line corresponds to
$\Omega=100\;{\rm s}^{-1}$.
The corresponding lines for $\Omega=0.1\;{\rm s}^{-1}$ and
$\Omega=1\;{\rm s}^{-1}$ cannot be resolved
from the zero-frequency limit. In part (b), all the lines for
the above mentioned four different frequencies merge with the zero-frequency
result (solid line). Note that for the two-state Markovian counterpart of the
considered three-state model with the rates depicted in Fig. 2(b),
all  the lines for the  different frequencies
merge with the zero-frequency result in part (a). Likewise, in part (b),
the depicted line becomes then be multiplied with the factor $1/N(0)\approx
147.43$ to yield the corresponding {\it Markovian  two-state} result.
}
\label{Fig3}
\end{figure}

The temperature-dependence of the corresponding Markovian transition
rates are given in Fig. 2(a).
For this set of parameters, the  coefficient of variation of the closed
residence time (compound state) within the three-state Markovian description
is $C_{c'}\approx 17.14$ and the low-frequency SNR is suppressed by
the factor of $1/N(0)\approx 147.43$ as compared with the corresponding two-state
Markovian case with the same rates $k_c^{(0)}$ and $k_{o'}^{(0)}:=
\langle \tau_{c'}\rangle^{-1}$. Its
temperature-dependence is depicted in  Fig. 2(b).
The frequency prefactor $\nu_0$ has been taken  to be
$\nu_0=6.11\cdot 10^{12}\;{\rm s}^{-1}$. This value corresponds
approximately
to a gas-phase value of $k_BT/h$ at $T=20^o$C.
It clearly overestimates any effects that relate
to friction and
other  details determining the transmission coefficient $\kappa$ in condensed phases.
The correct estimation of this prefactor would
require a more elaborate theory of the Kramers type \cite{HTB90} rather than
Absolute Rate Theory used here.
Nevertheless, this ambiguity does not play role if to assume that
the frequency prefactor $\nu_0$ is one and the
same for all the rate parameters. This is so because a different value
of $\nu_0$ would result in one and the same (unknown)
systematic entropy correction for all
{\it metastable} states. The entropy {\it differences} between
the {\it metastable} states remain unchanged.

Our results for this choice of parameters are depicted in Fig. 3(a)  for the
spectral amplification $\eta$, and Fig. 3(b) for the
corresponding signal-to noise ratio
$R_{SN}$.
%%%%%%% addition %%%%%
In particular, Fig. 3(a) convincingly demonstrates  the possibility of Stochastic
Resonance within the range of physiological temperatures when $\epsilon(T)
\approx 0$, in accordance with Eq. (\ref{universal}). The SNR in Fig. 3(b) also does
exhibit the SR-like increase with increasing temperature. It does not, however, reach a resonance
peak-behavior in the corresponding temperature range. This peak behavior would formally be assumed
(not depicted with Fig. 3(b)) only at  physiologically unrealistic
high temperatures of about 600 $^o{\rm C}$. This resonance behavior is formally contained in
Eq. (\ref{SNR0}) (notice the presence of $\nu(T)$ in the numerator there, which
invalidates the criterion $\epsilon(T)\approx 0$). The behavior at extreme
high frequencies $\Omega\gg k_{in}, k_r, k_o^{(0)}$ (cf. Eq. (\ref{SNR2}))
assumes qualitatively the same behavior as depicted  in Fig. 3(b); the only difference being
that the SNR becomes increased by the factor $(C_{c'}^2+1)/2$. Surprisingly (in view of the rather large
enthalpic barriers), this does not exclude  the appearance
of Stochastic Resonance in the SNR-behavior at physiological temperatures
at some intermediately high frequency of the signal. For the studied
model system this occurs, for example, for $\Omega=10^4\;{\rm s}^{-1}$
(this feature is not depicted because the accompanying spectral amplification $\eta(T)$ in Fig. 3(a)
emerges to be extremely small).

\subsection{The case with a power law distribution}

To model  the gating dynamics more realistically we next consider  a description with a nontrivial
nonexponential residence time distribution in the inactivated
regime $\psi(\tau)$ in
Eq.(\ref{Gfin}). In particular, we use
a probability density
$\psi(\tau)$ with the following characteristic function \cite{GH04}
\begin{eqnarray}\label{modelpsi}
\tilde \psi(s)= \frac{1}{1 + s\langle \tau_r\rangle
g_{\alpha}(s\tau_d)}\;,
\end{eqnarray}
where
\begin{eqnarray}
g_{\alpha}(z)=\frac{ \tanh(z^{\alpha/2})}{z^{\alpha/2}}.
\end{eqnarray}
The RTD (\ref{modelpsi}) has been obtained in Ref.\cite{GH04} as
solution of a conformational diffusion model accounting
for an anomalous subdiffusion over energetically quasi-degenerate
substates within the given conformation, i.e. $I$ in the present work.
This probability density possesses
three parameters: the mean residence time
$\langle \tau_r\rangle$ in this inactivated conformation, the conformational diffusion
time $\tau_d$ and the index of subdiffusion $\alpha$, $0<\alpha\leq 1$.

In spite of possessing three independent parameters only, the RTD (\ref{modelpsi}) is capable
to display a  rich behavior. The case of single-exponential distribution
with the rate parameter $k_r:=\langle \tau_r\rangle^{-1}$ is rendered for
$\tau_d=0$.
The case $\alpha=1$ corresponds to  normal diffusion.
For $\tau_d \ll
\langle \tau_r\rangle$, the intra-conformational diffusion effects are
not essential. However, for $\tau_d \gg \langle \tau_r\rangle$ and in the
range
$\langle \tau_r\rangle^2/\tau_d\ll\tau\ll \tau_d$
the corresponding RTD assumes a negative power law, i.e.,
$\psi(\tau)\propto \tau^{-3/2}$, which
ends up in an exponential tail for $\tau>\tau_d$ \cite{PNAS02}.
For $\alpha <1$,
the distribution (\ref{modelpsi}) has an infinite variance, since
$\langle \tau^2\rangle =\infty$ for
any $\tau_d \neq 0$ and, depending on a subtle interplay of parameters, it
can display up to three different power laws. These are:
$\psi(\tau)\propto \tau^{-\alpha/2}$ initially,
$\psi(\tau)\propto \tau^{-(2-\alpha/2)}$ intermediately, and
$\psi(\tau)\propto \tau^{-(2+\alpha)}$ asymptotically \cite{GH04}.
With the discussed choice of $\psi(\tau)$, the characteristic function
of the RTD in the compound $C'$ state reads
\begin{eqnarray}\label{compound4}
\tilde \psi_{c'}(s)= \frac{k_o^{(0)}(sg_{\alpha}(s\tau_d)+k_r)}{(s+k_o^{(0)})
[sg_{\alpha}(s\tau_d)+k_r ]+
k_{in}sg_{\alpha}(s\tau_d)}\;.
\end{eqnarray}
It possesses the same average residence time (\ref{meanRTD1}) and the coefficient
of variation is
\begin{equation}\label{cvpareto}
C_{c'}=\left\{ \begin{array}{r@{\quad \quad}l}
\infty, \alpha < 1;\;\;\\
 \sqrt{1+\frac{2k_o^{(0)}k_{in}}{(k_{in}+k_r)^2}(1+k_r\tau_d/3)}, \; \alpha = 1\;.
\end{array} \right.
\end{equation}
Note that for $\alpha<1$, $C_{c'}=\infty$. This implies that the detection of low-frequency
signals $\Omega\to 0$ is fully suppressed as $R_{SN}(\Omega)\to 0$ at $
\Omega\to 0$ \cite{PRL03,PRE04}. A bistable stochastic element with
such properties can thus be used
as a high-pass filter for the signal
transduction. In this case, the first leading terms
of the small-$s$ expansion of $\tilde \psi_{c'}(s)$ read
$\tilde \psi_{c'}(s)\approx 1-\langle \tau_{c'} \rangle s+ \frac{1}{3}
\frac{k_{in}}{k_o^{(0)}k_r\tau_d}(s\tau_d)^{1+\alpha}$. This corresponds
asymptotically to a distribution,
$\psi_{c'}(\tau)\propto \tau^{-(2+\alpha)}$, similar to the Pareto law behavior
considered in Refs. \cite{PRL03,PRE04}.

\begin{figure}
\epsfig{file=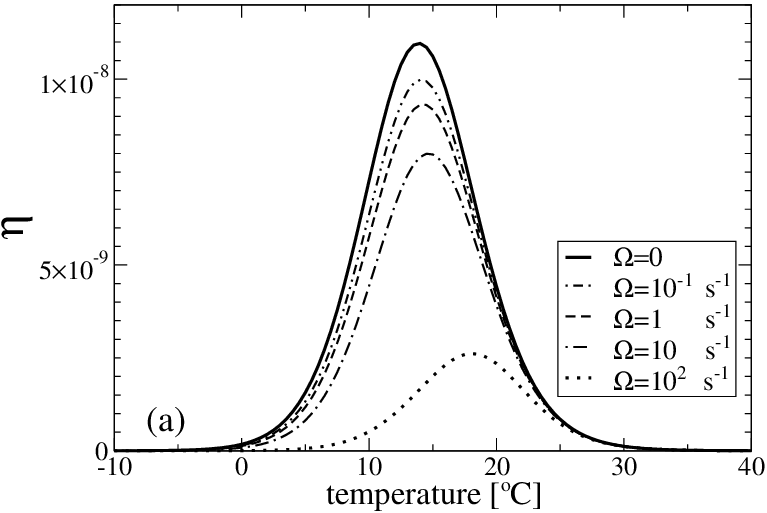,width=.5\textwidth}

\vspace{1cm}

\epsfig{file=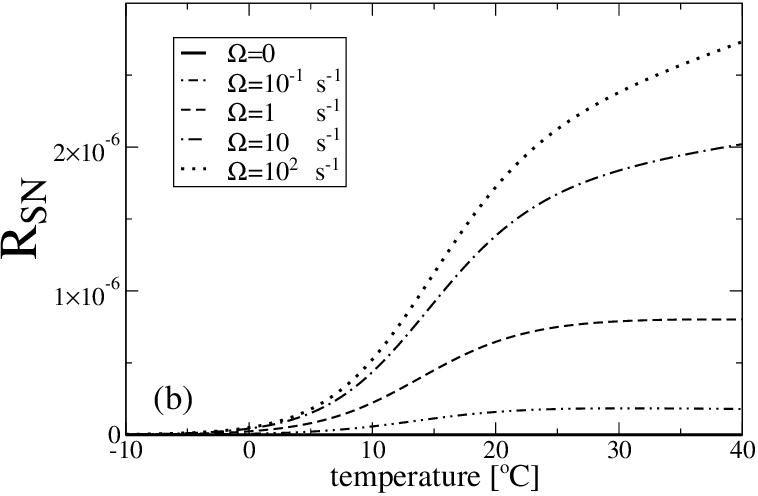,width=.5\textwidth}

\vspace{1cm}
\caption{The case with a power law distribution of  residence times
for  three-state ion channel gating, cf. Fig 1.
Instead of a single-exponential RTD $\psi(\tau)$,
the residence time  distribution in the inactivated state  
is given now by Eq.
(\ref{modelpsi})  with $\alpha=0.25$ and $\tau_d=0.1\; {\rm s}$. The other parameters
remain the same as in Fig. 3. The manifest non-Markovian
effects  result in a resolution of the different lines that merged in Fig. 3. Note  in part (b) that the
signal to noise ratio is fully suppressed towards zero  for $\Omega\to 0$ (the result merges with the
horizontal axis). The
adiabatic limit of the corresponding  Markovian three-state model
 is approximately assumed at
$\Omega=100\;{\rm s}^{-1}$; cf. the dotted line in (b) which compares with the solid line in
Fig. 3(b).}

\label{Fig4}
\end{figure}

With (\ref{modelpsi}) in Eqs. (\ref{Gfin}),
(\ref{eta}) and (\ref{res3}) one can evaluate the spectral
amplification of signal and the signal-to-noise ratio. We did this
for the above set of thermodynamical parameters (see in Table \ref{table})
entering rates which is
supplemented
with the following parameters of the RTD in Eq. (\ref{modelpsi}):
$\alpha=0.25$,
$\tau_d=0.1\,{\rm s}$.
As it can be seen in Fig. 4(a), the non-Markovian effects resolve
the lines at the frequencies
$\Omega =0.1\;{\rm s^{-1}}$ and $\Omega =1\;{\rm s^{-1}}$
which merge with the zero-frequency line in Fig. 3(a). Moreover,
the different lines which merge in Fig. 3(b) become also resolved,
see in Fig.4(b), where upon increasing the
angular frequency $\Omega$
the SNR $R_{SN}(\Omega,T)$ grows. Namely, at zero frequency
$R_{SN}(\Omega\to 0,T)=0$
and
for $\Omega\approx 100\,{\rm s^{-1}}$  $R_{SN}(\Omega,T)$ reaches
approximately (from below)
its low-frequency three state Markovian limit in Fig. 3(b). With the further
increase of $\Omega$ SNR will, however, grow further approaching asymptotically
the two state Markovian limit, where the non-Markovian form factor
assumes unity, i.e. $N(\Omega)=1$.
%%%%%% added %%%%
It is worth  noticing that for the two lower curves in Fig. 4(b) the SNR
seemingly  saturates with increasing temperature. In fact,  however, the SR-behavior
exhibits a rather  broad maximum. The occurrence of this maximum is quite surprising
(see the discussion at the end of the subsection IV.A) and is
due to non-Markovian effects.

Moreover, at variance with the SNR behavior (where the SNR increases with increasing angular  frequency
of signal) the corresponding SPA diminishes with
increasing  angular frequency. Therefore, we do not observe an intermediate frequency
regime which would prove optimal for the detection of non-Markovian SR.

\section{Resume}

With this work we have investigated Stochastic Resonance in a three
state non-Markovian model of ion channel gating. We note that our
scheme of a three-state non-Markovian modeling is distinctly
different from a similar one,  recently applied to an excitable
neuronal dynamics \cite{LGNS,Prager}. The latter model assumes a three
state system which cycles unidirectionally between three states, $1$
(silent), $2$ (excited) and $3$ (refractory), i.e., $1\to 2\to 3\to
1$. This unidirectional cycling corresponds to processes that are
very far from the thermal equilibrium and which require a continuous
supply of free energy.  In contrast, our modeling is compatible
with the thermal equilibrium. The ion channel gating is commonly
assumed to be a thermal equilibrium process \cite{Hille}. This is in
contrast to the situation with  ion pumps and neuronal systems
which do require a free
energy supply for proper functioning. As it has been shown in this
work, our three state description is compatible with the
phenomenological two-state theory of non-Markovian Stochastic
Resonance put forward in Ref. \cite{PRL03}, while the
non-equilibrium, three-state model of Ref. \cite{Prager} is not
within this class of system behaviors \cite{PRE04}. Nevertheless,
the result of Ref. \cite{Prager} for the spectral power
amplification can be reproduced from our Eq. (\ref{mainres}) by
specifying the RTD of the compound silent state (i.e. the silent
state plus the refractory state)  as a time-convolution of two
corresponding residence time probability densities and expanding it
like in Eq. (\ref{series}). In other words, the model of Ref. \cite{Prager}
can be recovered
as a special case within our general nonequilibrium approach put
forward in Ref. \cite{PRE04}, cf. section IV therein.

In conclusion, the  non-Markovian SR effects such as the suppression
of the signal-to-noise ratio for low-frequency signals and  the
resolution of different signal frequencies in the spectral
amplification of signal in the presence of possibly  large
(entropic) asymmetries can be nicely modeled already within a three
state Markovian model. Such a three state Markovian model yields the
simplest non-Markovian model after the corresponding projection onto
the subspace of two observable states. The signal-to-noise ratio
remains, however, finite in the limit of the zero-frequency signal.
Its complete suppression requires an infinite variance of the
residence time intervals in the inactivated state. For such a
manifest non-Markovian SR-behavior, weak external oscillating
signals with an intermediate frequency should be used in order to
detect SR experimentally.

%%%%% add %%%%
Our present results provide a theoretical proof for the occurrence of
Stochastic Resonance in {\it single} biomolecules at
physiological temperatures. This being so, our findings can guide
the experimentalists to choose both, the appropriate molecules for
doing experiments and to identify the corresponding experimental parameter regime
which in turn will reveal the SR-phenomenon in a single ion channel.  Presently there
exist only very few experimental, ion channel-based SR-works:  In Ref. \cite{Petracchi} attempts to identify
SR in a single ion channel have been made; the
occurrence of SR therein is, however, not  convincing. In  Ref. \cite{Bezrukov}
one investigates SR and demonstrates SR. The phenomenon has been studied, however, only on the
level of a small number of 
dynamically self-assembled alamethicin ion channels.

Our findings show that ideal candidates to observe 
experimentally SR on the level of {\it single} molecules are cold (or heat) sensitive
ion channels \cite{Nature04}. From this viewpoint, we believe that a repetition of the
experiments in  Ref. \cite{Petracchi} by resorting to
such channels would indeed become successful for observing SR
in a single ion channel operating within its physiological regime.

\section{acknowledgements}

This work has been in part supported by the Deutsche Forschungsgemeinschaft
within SFB-486 ``Manipulation of matter on the nanoscale'', project
A10, and by the Ministry of Science and Education (Spain),
project FIS2004-02461.
JLV would like to thank all members of the Institute of Physics of the
University of Augsburg for their kind hospitality during his stay in Germany.

\appendix

\section{\label{append}Derivation of linear response function within
non-Markovian two-state theory}

In this Appendix the result in Eq. (\ref{PHLR}) is re-derived from the
non-Markovian two-state theory of Ref. \cite{PRE04}.
Namely, in Ref. \cite{PRE04} it has been shown that the linear
response function of an arbitrary two state renewal process characterized
by the conditional RTDs $\psi_{j=1,2}(\tau|t)$ and the amplitude of conductance fluctuations
$\Delta g$ to a periodic
signal in Eq. (\ref{signal}) is given in the limit $A\to 0$ by
\begin{eqnarray}\label{mainres}
\tilde \chi(\Omega)=-\frac{2i\Delta g}{A\Omega}\frac{1}{\langle \tau_1\rangle
+\langle \tau_2\rangle}\frac{\tilde \psi_2^{(1)}(-i\Omega)
[1-\tilde \psi_1(-i\Omega)]-\tilde \psi_1^{(1)}(-i\Omega)
[1-\tilde \psi_2(-i\Omega)]}{1-\tilde\psi_1(-i\Omega)\tilde\psi_2(-i\Omega)}.
\end{eqnarray}
In Eq. (\ref{mainres}), $\tilde \psi_{j}(s)$ denote the Laplace-transforms of
stationary the RTDs in the absence of driving and $\tilde \psi_{j}^{(1)}(s)$
are the Laplace-transforms of the corresponding contributions in the expansion of the
conditional, driven RTDs, i.e.,
\begin{eqnarray}\label{series}
\psi_j(\tau|t)=\sum_{n=-\infty}^{\infty} \psi_j^{(n)}(\tau)\exp(-in\Omega t).
\end{eqnarray}
The quantities $\tilde \psi_{j}^{(1)}(s)$ must be evaluated from an underlying multi-state, or
continuous state dynamics to first  order in $A$. Whenever $\tilde \psi_{j}^{(1)}(s)$
satisfy the relation
\begin{eqnarray}\label{link}
\tilde \psi_{j}^{(1)}(-i\Omega)=-\frac{1}{2}\beta_{j} A [1-\psi_j(-i\Omega)],
\end{eqnarray}
then the result of phenomenological theory in Eq. (\ref{PHLR}) is recovered from
Eq. (\ref{mainres}). For the exponential form of distribution of open residence
times $\psi_o(\tau|t)=\psi_2(\tau|t)=k_c(t+\tau)\exp[-\int_t^{t+\tau}k_c(\tau')
d\tau']$, the relation (\ref{link}) is fulfilled. We identify here the state ``1''
with the compound closed state ``(C, I)'' and the state ``2'' with the
open state ``O'';  note also
 that $\beta_1\equiv \beta_o$, $\beta_2\equiv\beta_c$.
This relation is valid as well for a
multi-exponential distribution which assumes a scaling relation among the rate
parameters which is not modified by driving, i.e. a form-invariant RTD, cf.
\cite{PRE04}. Below we demonstrate that the relation (\ref{link}) is also valid in the
present case which obviously
does not belong to the latter universality class. Nevertheless, the application of
the phenomenological
linear response theory to the considered non-Markovian processes \cite{PRL03}
is justified.

In order to obtain the conditional survival probability of the compound
closed state $\Phi_1(\tau|t):=\Phi_1(t+\tau,t)$ one needs to solve the system
of equations
\begin{eqnarray}\label{cond}
 \frac{d}{d\tau}p_c(\tau|t)& = &-[k_{in}+k_o(t+\tau)]p_c(\tau|t)+\int_{0}^{\tau}
\Gamma(\tau-\tau')p_I(\tau'|t)d\tau'\;, \nonumber \\
\frac{d}{d\tau} p_I(\tau|t) & = &k_{in} p_c(\tau|t) - \int_{0}^{\tau}
\Gamma(\tau-\tau')p_I(\tau'|t)d\tau'\;,
\end{eqnarray}
with the initial conditions $p_c(0|t)=1$ and $p_I(0|t)=0$. Then,
$\Phi_1(\tau|t)= p_c(\tau|t) + p_I(\tau|t)$.
Asymptotically, the considered periodically driven renewal process becomes
cyclic-stationary.
For such a cyclic-stationary process,
$\Phi_1(\tau|t)$ must be invariant under time shifts $t\to t+ nT,\; n=\pm 1,\pm 2,...$
with period $T=2\pi/\Omega$; it thus  can be expanded
into Fourier series like in (\ref{series}). With $k_o(t)$ being a periodic function
of time, we are seeking a solution of
Eq. (\ref{cond}) in the form, $p_\alpha(\tau|t)=\sum_{n=-\infty}^{\infty}
p_\alpha^{(n)}(\tau)\exp(-in\Omega t), \; p_\alpha^{(-n)}(\tau) =
[p_\alpha^{(n)}(\tau)]^*, \; \alpha=c,I$. Invoking additionally
a small-signal expansion (\ref{small-rate}) this yields an infinite
system of coupled integro-differential equations for
$p_\alpha^{(n)}(\tau)$ which upon the
use of the Laplace-transform method results in
a recurrent-difference relation for $\tilde p_c^{(n)}(s)$, i.e.,
\begin{eqnarray}\label{R}
\Big [s + k_{in}+ k_o^{(0)}-\frac{k_{in}\tilde \Gamma(s)}{s+\tilde \Gamma(s)}
\Big ]\tilde p_c^{(n)}(s) +
k_o^{(1)}[\tilde p_c^{(n+1)}(s-i\Omega) + \tilde p_c^{(n-1)}(s+i\Omega)]=
\delta_{n,0}\;,
\end{eqnarray}
and a relation expressing $\tilde p_I^{(n)}(s)$ through $\tilde p_c^{(n)}(s)$, reading,
\begin{eqnarray}\label{A5}
\tilde p_I^{(n)}(s)=\frac{k_{in}}{s+\tilde \Gamma(s)}\tilde p_c^{(n)}(s),
\end{eqnarray}
where $k_o^{(1)}\propto A$ is given in Eq. (\ref{k1}). Note that $\tilde
p_\alpha^{(n)}(s)$ are nonlinear functions of the driving amplitude $A$. The
equation (\ref{R}) can be solved perturbatively by using the corresponding expansions
of $\tilde p_\alpha^{(n)}(s)$ in $A$.
The expansion of $p_\alpha^{(0)}(s)$ in $A$ starts from the driving independent
terms and obviously (from the symmetry reasons, cf. $A\to -A$) does not contain
linear contribution in the amplitude strength $A$.
Moreover, $\tilde p_\alpha^{(\pm 1)}(s)\propto A$,
to the  lowest order in $A$. Hence, $\tilde p_\alpha^{(\pm 2)}(s)=O(A^2)$
because the response at the second harmonic driving frequency cannot be linear.
The use of standard perturbation theory then yields to the lowest order in $A$,
\begin{eqnarray}
\tilde p_c^{(0)}(s) & = &\frac{1}{s+k_{in}+k_o^{(0)}-\frac{k_{in}\tilde \Gamma(s)}{s+
\tilde \Gamma(s)}} + O(A^2)\;, \label{A7}\\
\tilde p_c^{(1)}(s) & = -&\frac{k_o^{(1)}\tilde p_c^{(0)}(s+i\Omega)}{s+k_{in}+k_o^{(0)}-\frac{k_{in}\tilde \Gamma(s)}{s+
\tilde \Gamma(s)}}\;.\label{A8}
\end{eqnarray}
Eq. (\ref{A7}) determines in combination with Eq. (\ref{A5}) for $n=0$ the
stationary (i.e., in the absence of  driving) survival function of the compound closed
state in Eq. (\ref{compound0}) and the corresponding RTD (\ref{compound1}), respectively.

Eq.(\ref{A8}) yields in virtue of  $\tilde p_c^{(0)}(0)=1/k_o^{(0)}+O(A^2)$
and Eq. (\ref{k1}) the result
\begin{eqnarray}
\tilde p_c^{(1)}(-i\Omega) & = &\frac{1}{2} A\frac{\beta_o}
{-i\Omega+k_{in}+k_o^{(0)}-\frac{k_{in}\tilde \Gamma(-i\Omega)}{-i\Omega+
\tilde \Gamma(-i\Omega)}}.
\end{eqnarray}
Together with Eq. (\ref{A5}) for $n=1$ this gives with $\beta_o=\beta_1$
\begin{eqnarray}\label{last}
\tilde \Phi_1^{(1)}(-i\Omega)=\frac{1}{2}A\beta_1 \tilde \Phi_1^{(0)}(-i\Omega).
\end{eqnarray}
Upon using the general relation, $\tilde \psi_\alpha^{(n)}(s)=\delta_{n,0}-
s \tilde \Phi_\alpha^{(n)}(s)$, Eq. (\ref{last}) thus yields the desired relation in (\ref{link}).

\end{document}